
 \magnification=\magstep1
\settabs 18 \columns
\hsize=16truecm

\def\s{\sigma}

\def\b{\bigskip}
\def\bb{\bigskip\bigskip}

\def\no{\noindent}
\def\r{\rightline}
\def\ce{\centerline}
\def\ve{\vfill\eject}

\def\r{\rightline}
 \font\got=eufm8 scaled\magstep1
 \def\g{{\got g}}

\def\s{\sigma}

\def\harr#1#2{\smash{\mathop{\hbox to .25 in{\rightarrowfill}}
 \limits^{\scriptstyle#1}_{\scriptstyle#2}}}

\def\today{\ifcase\month\or January\or February\or March\or April\or
May\or June\or July\or
August\or September\or October\or November\or  December\fi
\space\number\day, \number\year }

\r \today
\bb\bb\bb


\def\Rrm{\hbox{\rm I\hskip -2pt R}}

\def\sqr#1#2{{\vcenter{\vbox{\hrule height.#2pt
\hbox{\vrule width.#2pt height#2pt \kern#2pt
\vrule width.#2pt}
\hrule height.#2pt}}}}

 \def\1/2{{\scriptstyle{1\over 2}}}
 \def\a/2{{\scriptstyle{3\over 2}}}
 \def\5/2{{\scriptstyle{5\over 2}}}
 \def\7/2{{\scriptstyle{7\over 2}}}
 \def\3/4{{\scriptstyle{3\over 4}}}

\def\picture #1 by #2 (#3){
  \vbox to #2{
    \hrule width #1 height 0pt depth 0pt
    \vfill
    \special{picture #3} 
    }
  }

\def\scaledpicture #1 by #2 (#3 scaled #4){{
  \dimen0=#1 \dimen1=#2
  \divide\dimen0 by 1000 \multiply\dimen0 by #4
  \divide\dimen1 by 1000 \multiply\dimen1 by #4
  \picture \dimen0 by \dimen1 (#3 scaled #4)}
  }

%
%

\font\steptwo=cmb10 scaled\magstep2
\font\stepthree=cmb10 scaled\magstep3
\magnification=\magstep1

  {\ce {\stepthree   Reissner-Nordstrom and charged gas spheres.}}  
\b
 \ce{Christian Fr\o nsdal}
\b
 \ce{\it Physics Department, University of California, Los Angeles CA
 90095-1547 USA}
\vskip1in

\def\sqr#1#2{{\vcenter{\vbox{\hrule height.#2pt
\hbox{\vrule width.#2pt height#2pt \kern#2pt
\vrule width.#2pt}
\hrule height.#2pt}}}}

\def \r{\rightarrow}

\def\e{ \rm e} 

\no{\it ABSTRACT}.  The main point of this paper is a suggestion about the
proper treatment of the photon gas in a theory of stellar structure
and other plasmas. This problem arises in the study of
polytropic gas spheres, where we have already introduced some innovations.
The main idea, already advanced in the context
of neutral, homogeneous, polytropic stellar models, is to base the theory
firmly on a variational principle. Another essential novelty is to let
mass distribution extend to infinity, the boundary between bulk and
atmosphere being defined by an abrupt change in the polytropic index,
triggered by the density.  The logical next step in this program is to
include the effect of radiation, which is a very significant complication
since a full treatment would have to include an account of ionization,
thus fields representing electrons, ions, photons, gravitons and
 neutral atoms as well. In way of preparation, we
consider models that are charged  but homogeneous, involving only
gravity, electromagnetism and a single scalar field that represents
both the mass and the electric charge; in short,  a
non-neutral plasma. While this work only represents a stage in the
development of a theory of stars, without direct application to
physical systems, it does shed some light on the meaning of the
Reissner-Nordstrom solution of the modified  Einstein-Maxwell
equations., with an application to a simple system.

\b\b

\no{\steptwo 1. Introduction}

A  fundamental field theory is characterized by the fact that the
number of degrees of freedom   is specified from the start, usually
revealed by counting the independent field components that appear in
the action. The standard approach to stellar structure is very
different. First, there is the mysterious vector field that is said to
represent the flow; it  is usually normalized to unity, a step that we
find very hard to accept. In the study of spherically symmetric,
equilibrium configurations it contributes at most one degree of
freedom, in evolution, up to three, depending on the symmetry. Next
there are two scalar fields identified as energy (or mass) density and
pressure density. Additional fields are introduced as needed, including
additional densities, pressures, luminosity. Typically, the
introduction of each new field is accompanied by one or more
additional equations that help  to pin it down. (Surprisingly, 
the density of the photon gas is not dignified by the introduction of
an independent degree of freedom.) This type of approach requires a
tremendous amount of physical intuition, far exceeding the understanding
of this author. And he misses the guidance that comes with an action
principle.

A preliminary study of homogeneous, polytropic models of ideal stars [F1]
has shown that the incorporation of an action principle leads to 
a more coherent theory. The mass was shown to be a constant
of the motion and related to a conserved current.  The study was
limited to irrotational flows, and this limitation will be retained in
the present paper.  The action includes, besides the Hilbert action for
the metric field,  a straightforward, relativistic generalization of the
standard hydrodynamic action for irrotational motion [FW], namely
$$
A_{Matter}\propto\int dx\sqrt{-g}  \Bigg({ \rho\over
2}(g^{\mu\nu}\psi_{,\mu}\psi_{,\nu}-c^2) - V[\rho]\Bigg)\,.\eqno(1.1)
$$
It involves a single scalar field $\rho$  and
the velocity potential
$\psi$ with the dimension of length. A more realistic
treatment of stars such as the sun must take radiation into account.
The next logical step is therefore to invent an action principle that
involves charged matter as well as electromagnetic and gravitational
fields. To be realistic, one would have to include free electrons, ions,
and atoms. Here we shall deal with a simpler situation, not applicable
to the sun, but perhaps to certain layers of some stellar atmospheres,
and other non neutral plasmas.

It may be possible to account  for the effect of radiation by simply
adding the radiation pressure to the pressure of the matter component, 
as may seem to be justified by the fact that the value  taken by the
Lagrangian density on shell can be interpreted as pressure.  The
phenomenological device of taking them to be proportional to each other
[E][C] may also be justified. It is expected that the theory will
enlighten us on this point. 

The simplest expansion of the model to
include electromagnetism utilizes the fact that it has a conserved
current, 
$$
J^\mu = \sqrt{-g}\,\rho \,g^{\mu\nu}\psi_{,\nu}.\eqno(1.2)
$$
We shall proceed by
introducing the minimal electromagnetic interaction associated with this
current. In addition, we could just add the Maxwell action,
$$
A_{Maxwell} = {-1\over 16\pi}\int dx  \sqrt{-g} 
g^{\mu\nu}g^{\lambda\rho} F_{\mu\lambda}F_{\nu\rho} .
$$ 
But this is probably not the right thing to do, as we shall try to
show.

The simplest of all couplings of the metric to matter would use the
action for a neutral scalar field, as (1.1) but with $\rho = 1$ and $V=
0$. This is wholly inadequate for dealing with a
continuous matter distribution described  by macroscopic density and
pressure. To generate a pressure one has to include interactions, and
as it turns out the inclusion of the density $\rho$ and the potential 
$V[\rho]$ is an effective way to do that. The
Maxwell action is inappropriate for the same reason; we wish to
include  the effect of radiation, an incoherent superposition of a
large number of photons, endowed as the matter distribution with a
pressure and a density of its own. The traditional approach is to add
the radiation pressure, by hand, to the pressure of matter, without
attempting to include it as an additional variable in the dynamical
scheme. What appears to be needed is a contribution to the action of the
form
$$
A_{Radiation} = \int dx\sqrt{-g} \big( {-  \sigma \over 16\pi}
g^{\mu\nu}g^{\lambda\rho} F_{\mu\lambda}F_{\nu\rho} -W[\sigma] 
 \big ).
\eqno(1.3)
$$
Here $\sigma$ is a dimensionless scalar field and $W[\sigma]$ is an
internal energy density. Here too, it is expected that the introduction
of a density and a potential is an effective way to introduce the
interactions that give rise to a photon pressure. 

Inevitably, this leads to the idea that the gravitational radiation that
must be present has to be treated analogously, but that will not be
developed in this paper. (See [MM].)

This study will be further limited to systems with spherical symmetry
and to potentials of the form $V[\rho] \propto \rho^\gamma$, leading to a
polytropic equation of state.
\b
\b

(Added in response to referee. To be inserted here. See response to
referee.)
\b

The idea that the Reissner-Nordstrom metric can be interpreted in terms
of a charged gas was already advanced by Felipe {\it et al} [dF1][dF2].

\b

(End of addition.)

\bb

\ce{\bf Outline}

Section 2 is a somewhat slowly paced introduction to explain our point of
view. Section 3 is a study of static configurations of charged
polytropes, with an investigation of what we take to be the correct
physical interpretation of the Reissner-Nordstrom metric.   Among the
equations of motion we emphasize (1) the conservation law for the
current, (2) variation of the mass density $\rho$ gives a relation
between potentials that, in the limit when $\rho$ is zero, leads
directly to the extremal Reissner-Nordstrom metric, and (3) variation
with respect to the photon density $\sigma$ gives an  
expression for   $\sigma$ in terms of the field strength.

To complete the definition of the model we take
$$
V[\rho] = a \rho^\gamma,
$$
with $a$ and $\gamma$ piecewise constant. This gives rise to an
polytropic equation of state with index $n$,
$$
p = {a\over n} \rho^\gamma, ~~ \gamma = 1 + {1\over n}.
$$
It is well known that a value $n < 5$ in the bulk of the star is
compatible with regularity at the center. At great distances it is
customary to take
$\rho = 0$, certainly an excellent approximation, but another
polytrope, with $n > 5$ is also possible. The boundary presents
difficult problems in either case, a difficulty that we circumvent by
having the change in the index triggered by the density. Boundary
conditions are imposed only at the center and at infinity and all the
fields are continuous at the `boundary'. The photon internal energy
$W[\sigma]$ is chosen so as to give the usual equation of state, $ p_{\rm
ph} =  T_0^0/3$ for the photon pressure $ p_{\rm ph}$.
 
The theory has an interesting connection to the Reissner-Nordstrom
solution of Einstein's field equations for empty space, and the
extremal solution (where the electrostatic repulsion exactly cancels
the gravitational attraction) plays a role.

Direct applications to real physical systems are not discussed. The main
result of numerical calculations is that the photon density profile  
may be very different from the matter distribution. If this carries over
to the case of neutral plasmas then it will instill some skepticism with
respect to the assumptions that are usually made about the radiation
pressure. 

A discussion of lessons learned from this investigation is deferred to
the last section.

\ve

 \no{\steptwo 2. Background on method} 
\b
{\ce{\bf 2.1. Geodesics}}
A very attractive feature of General Relativity is the fact that a test
particle in a given metric field moves along a geodesic, minimizing the
action
$$
A_{Particle} \propto \int_\gamma ds,~~ ds^2 = g_{\mu\nu}dx^\mu dx^\nu.
$$
It is a well defined line integral along a path $\gamma$ with fixed
endpoints; it is independent of any parameterization of $\gamma$. If
$\tau:\gamma\rightarrow \Rrm$ is a parameterization, then
$$
A_{Particle} \propto \int d\tau\sqrt{g_{\mu\nu}U^\mu U^\nu},~~U^\mu =
dx^\mu/d\tau.
$$
The velocity of the particle at any point on the path is
$$
\vec v = d\vec x/dt,~~ d\vec x := (dx^1,dx^2,dx^3);
$$ 
it has a direct physical meaning, while the 4-velocity field $U$ 
does not. Since $\vec v = \vec U/U^0$, the vector field $U$ is defined up
to a scale transformation. The scalar field $g_{\mu\nu}U^\mu U^\nu$ is a
constant of the motion and that allows fixing its value along the  path,
therefore the scale may be fixed by requiring, for example, by setting
$ 
g_{\mu\nu}U^\mu U^\nu = 1. 
$ 
By referring to a ``test particle" one implies that the reaction of the
metric field to the presence of the particle is being neglected. In the
case of several test particles one is also neglecting any interaction
between the particles; in this case each particle has its own path, its
own 4-velocity field and its own ``proper time". It is possible to
fix the fourth, unphysical component (equivalently, the scale) of each, by
setting $g_{\mu\nu}U_1^\mu U_1^\nu =  g_{\mu\nu}U_2^\mu U_2^\nu = ... =
1$, but as the number of particles grows the neglect of any reaction on
the field, and of any interaction between the particles, makes such
conditions increasingly implausible.
\b
\ce{\bf 2.1. Field equations}

Einstein's field equations deal, in the first instance, with the
determination of the metric field produced by a given matter
distribution,
$$
G_{\mu\nu} = {8\pi G\over c^2} T_{\mu\nu},\eqno(2.1)
$$
where   $G_{\mu\nu} = R_{\mu\nu} -
(1/2)g_{\mu\nu}R$ and $T$ is the energy momentum tensor of the matter
distribution. All the familiar, elementary, covariant field theories
possess energy momentum tensors with the requisite properties, most
important of which is that of being covariantly conserved,
$$
 {T^{~\nu}_{\mu;\nu}}   = 0,\eqno(2.2)
$$
as it must be because of the (contracted) Bianchi identities that
are satisfied identically by the Einstein tensor. It is a direct
consequence of the fact that these theories are characterized by an
invariant action principle, with some action $A$ and
$$
T_{\mu\nu} = 2{\delta A\over\delta g^{\mu \nu}}.\eqno(2.3)
$$

In general, the right hand side of Eq.(2.1) is supposed to represent
`matter', the source of the metric field. It is plausible, and it seems
almost mandatory, that it have the form (2.3). But in studies of stellar
structure `matter' is extremely complicated and explicit, realistic
expressions for the action, in terms of elementary particle fields, is not
to be dreamt of.

It is necessary  to introduce statistical or more precisely
thermodynamical variables. If one excludes crystalline structures, then
it is natural to follow Tolman [T], who assumed that $T$ have the
representation
$$
T_{\mu\nu} = c^{-2}(\hat \rho + \hat p)U_\mu U_\nu -
\hat p\g_{\mu\nu},\eqno(2.4)
$$
where $\hat \rho$ and $\hat p$ are scalar fields and $U$ is a 4-vector
field that is associated with the local velocity of flow,
$$
\vec v(x) = \vec U(x)/U^0(x).\eqno(2.5)
$$
While the 4-velocity of a test particle is defined over the path, this
one is defined over the entire space time manifold, in accord with the
interpretation of the continuum as a limit of a dense distribution of
particles. Here too one has the problem of interpreting the fourth
component, and the remedy has always been to postulate the
``normalization"
$$
g(U) :=g_{\mu\nu}(x)U^\mu(x) U^\nu(x) = 1;\eqno(2.6)
$$
that is,   the highly implausible continuum limit of the
 normalization  used for $n$ non-interacting particles.

In defense of (2.6) it is true that any other normalization, $g(U)
= f^2$, say, with $f$ a scalar field, can be reduced to (2.6) by a
rescaling of $U ,\hat\rho$ and $\hat p$. This would affect the
physical interpretation of $\hat \rho$ and $\hat p$, and hence the
equation of state. In the case of slow motion and weak gravitational
fields Eq.(2.6) reduces to
$U^0(x) = 1$,  and in this case
$\hat\rho$ and $\hat p$ are interpreted as energy density and pressure,
respectively, to be treated as standard thermodynamic variables.
But the subsequent extrapolation of thermodynamics to velocities
approaching that of  light, and strong gravitational fields, is a bold
extension of the equivalence principle, making Eq.(2.6) an {\it ad hoc}
assumption with real physical consequences.

A more serious result of adopting Eq.(2.6) is that it cannot be
reconciled with an action principle. The required covariant conservation
law (2.2) thus becomes an axiom in itself, without an action principle to
support it. The power of action principles in the formulation of physical
theories is almost universally recognized, so one is not surprised to
find that attempts have been made to find an action consistent with
(2.6) [S][T], but  so far  these efforts have been formalistic and
without real content, which is why they have had no influence on
the development of the subject. 

In this paper we are beginning to work
with a several kinds of ``matter", each making a contribution to the
total action; in this case the formula (2.4) can no longer hold and the
question of normalization becomes moot.   In the case of a static
configuration it is still plausible to identify
$T_0^0$ with the total energy density. 
\ve
 
\ce{\bf 2.3. The Newtonian limit}

The equations of static, Newtonian gravity, a limiting case of
Einstein's theory, are Poisson's equation,
$$
\Delta \phi = 4\pi G\rho,\eqno(2.7)
$$
where $\rho$ is the mass density and $\phi$ is the gravitational
potential, and the hydrostatic condition
$$
\rho\vec\bigtriangledown \phi + \vec \bigtriangledown p = 0. \eqno(2.8)
$$

An important theory of irrotational hydrodynamics is based on Bernoulli's
equation. This theory can be formulated as a variational problem with the
action [FW]
$$
\int d^3x dt\Big(\rho(\dot \Phi - \vec v^2/2 -\phi) - V\Big ).\eqno(2.9) 
$$
Here $\Phi$ is the velocity potential, $\vec v = -\vec
\bigtriangledown\Phi,\, V$ is the internal energy and $\phi$ is the
external gravitational potential. In the isentropic case $V = V[\rho]$ is
determined by $\rho$ and the pressure is
$$
p = \rho{dV\over d\rho} - V.\eqno(2.10)
$$
This coincides (on-shell) with the action density, which explains why
pressure is additive. In this case the equations that govern equilibrium
configurations are Poisson's equation for $\phi$, the continuity equation
for the current $\rho\vec v$, and the variational equation
$$
\dot \Phi = \vec v^2/2 + \phi  + {dV\over d\rho}.\eqno(2.11)
$$
Taking the gradient of this equation one recovers, 
with the help of (2.10), the hydrostatic equation (2.8).

An effect of specifying an  action is thus to replace the differential,
hydrostatic condition by an integrated form of it, with a fixed choice of
the integration constant. In Newtonian gravity the zero point of the
potential has no meaning, so that this fixing of an integration constant
is irrelevant, but it has important consequences in General Relativity.
\b
 
\ce{\bf 2.4. Irrotational flow in General Relativty} 

A straightforward generalization of the action (2.9) is
$$
A_{Matter} =  \rho_{cr}\int d^4x\sqrt{-g}\Bigg({\rho\over 2}(g^{\mu\nu}
\psi_{,\mu}\psi_{,\nu} - c^2) - V[\rho]\Bigg),\eqno(2.12)
$$
where $\rho$ and $\psi$ are scalar fields. The field $\rho$ is
dimensionless; the correct density dimension is provided by the factor
$\rho_{cr}$.The non-relativistic theory is recovered by setting
$$
\dot \psi = c^2  + \dot\Phi,~~ \partial_i \psi = \partial_i\Phi,~~ i =
1,2,3,
\eqno(2.13) 
$$
or simply $\psi = c^2t + \Phi$.

We consider the theory to be defined by the action.  The possibility of a
thermodynamical interpretation, for some choice of 
the functional $V[\rho]$, is expected, but it is not an axiom. This is an
important departure from the traditional approach.

 It is natural to define
the material pressure as before, by Eq.(2.10),
$p =
\rho(dV/d\rho) - V$, a scalar field that, on the trajectory, coincides
with the Lagrangian density.

Variation of the independent fields $\psi$ and $ \rho$ leads to the
equation of continuity
$$
\partial_\mu J^\mu = 0,~~ J^\mu := \rho\sqrt{-g} g^{\mu\nu}
\psi_{,\nu}\eqno(2.14)
$$
and the analogue of  (2.11), namely
$$
{1\over 2} (g^{\mu\nu}
\psi_{,\mu}\psi_{,\nu} - c^2) = {dV\over d\rho}.\eqno(2.15)
$$

The energy momentum tensor associated with the action (2.12) is given
on shell  by
$$
T_{\mu\nu} =  \rho\psi_{,\mu}\psi_{ ,\nu} -
pg_{\mu\nu}.\eqno(2.16)
$$
It is of the Tolman form, except that the vector field $U$ is here
restricted to be of the gradient type. Eq.s (2.14-15) imply that 
${T^{~\nu}_{\mu;\nu}} = 0$. On the other hand, to prove it, one needs
only the gradient of (2.15), namely
$$
\partial_\lambda p = {\rho\over
2}\partial_\lambda (g^{\mu\nu}\psi_{,\mu}\psi_{,\nu}).\eqno(2.18)
$$
Thus one sees that here too, as in the non-relativistic theory, the
first effect of introducing an action is to fix an integration
constant.

\b
\ce{\bf 2.5. Implications of variational formulation for  boundary
conditions}

In any analysis of stellar structure the static solutions play a dominant
role. They describe the equilibria and, in addition, time
development is often adiabatic and hence a succession of equilibrium
configurations. It is important to notice that, in Tolman's theory, some
of the metric fields are represented only by their derivatives.  
We consider the case of static configurations in a metric that
embodies spherical symmetry and, in a set of coordinates
$t,r,\theta,\phi$, takes the form
$$
ds^2 = \e^\nu dt^2 - \e^\lambda dr^2 - r^2 d\Omega^2,\eqno(2.19)
$$
with $\nu,\lambda$ depending on $r$ only. The field $\lambda$ appears in
the standard field equations (Tolman's approach), as does
$\nu':=d\nu/dr$, but
$\nu$ itself does not. The equations are therefore insensitive to a
constant shift of $\nu$, for this reason the boundary conditions are less
restrictive than in the model based on the action principle.

An external, Schwarzschild metric has $\lambda+\nu = 0$; therefore, at
the surface of a star, where $r = R$, say, the following boundary
condition must hold,
$$
 \lambda(R) + \nu(R) = 0  .\eqno(2.20)
$$
Consider a homogeneous, polytropic model of the sun:
$$
V = a\rho^{4/3},~~ a {\rm ~constant}. 
$$
A determination of the best value of the constant $a$ would take us too
far, see [F3], so we eliminate it,
$$
{dV\over d\rho} = 4{p\over\rho},\eqno(2.21)
$$
then use the gas law to define the temperature,
$$
T = {\mu\over {\cal R}}{p\over \rho}.\eqno(2.22)
$$
In the static case $\dot\psi = 1,\, \vec\bigtriangledown\psi = 0$ and
Eq.(2.15) becomes
$$
-\phi = {1\over 2}({\rm e}^{-\nu} -1) = 4{p\over\rho} = 4{\mu\over {\cal
R}}T.\eqno(2.23)
$$
If $\rho$ vanishes for $r>R$ then matching the metric to the external
Schwarzschild metric at $r=R$ gives $\phi(R) = -GM/R$, where $M$ is the
mass. Thus finally,  
$$
T(R)= {1\over 4}{\mu\over {\cal
R}}{GM\over R} .
$$ 

The standard approach gives this temperature distribution (2.23) {\it
modulo} an additive constant; matching to the external
Schwarzschild metric is inconsequential. In this paper we shall
discover additional confirmation of the relevance of the stronger
condition (2.15).
\b
\ce{\bf 2.6. Stability, mass  and improved boundary conditions}
 
To complete the matter model we take
$$
V[\rho] = a\rho^\gamma,
$$ with $a$ and $\gamma$ piecewise constant. This gives rise to the
isentropic equation of state,
$$
p = {a\over n} \rho^{\gamma},~~ n =   {1\over \gamma-1},
$$
of a polytrope with index $n$. It is well known that taking $n<5$
in the bulk of the star ensures that the density becomes zero at some
finite radius.. 

We have reported calculations of the static solutions of this model,
for various values of the polytropic index [F3]. Rather than
follow  the traditional method of defining the boundary  
of the star to be at the first zero of the density, we matched the
interior solution to an exterior Schwarzschild metric.  This strategy
does not work in the traditional context, but the stronger
wave equations obtained from variational principle determine the
radius uniquely as the place where the condition (2.20) holds, the
only place where matching to the exterior metric is possible.

 Subsequently we
studied the stability of these static configurations [F3].  It turned
out that we could not find sufficient guidance to choose between the
possible boundary conditions. In particular, it seemed
strange that the mass, defined asymptotically by the exterior metric,
turned out to echo the oscillations of the star.   As a first attempt
to remedy the situation we replaced the exterior Schwarzschild metric
by another polytrope, with $n > 5$. Unfortunately, that was not enough 
to settle the question of appropriate boundary conditions. It was
decided to give up the idea of a fixed boundary alltogether. After
all, the `boundary' of a double polytrope is just a place where
the polytropic index changes, more or less abruptly. If the star
starts from a diffuse, gaseous state, then this change must come about
as a result of the increase in density that follows from the gradual
gravitational collapse and in response to the threat of a singularity
developing near the center.

 The change in the index at ``the boundary'' is evidently a
result of the increase in density, and all uncertainty concerning the
correct choice of boundary conditions can be avoided by making the
interdependence of index and density explicit, posing for   example,
$$
n = 3+ {  3  \over 1 +  \rho^K},
$$
where $K >> 1$ and $\rho =1$ is a critical density. The boundary
conditions at the center are  supplemented by the natural requirement
that the fields decrease at infinity. As it turns out, this radical
change in the treatment of the boundary has very little effect on the
static solutions, but a very salutary influence on stability. Boundary
conditions are imposed at the center and at infinity; all static
solutions found appear to be quite stable to radial excitations.
In addition, it turned out that the new boundary conditions (at
infinity) ensure that the mass, defined by the asymptotic metric, but
also related to the space integral of the charge density $\rho$ (with
the correct measure!) is a constant of the motion. This is regarded as
an important advantage over the traditional theory. The same strategy is
followed in the present paper.

\b
 \b

\no{\steptwo 3. Radiation and charged matter}

\ce{\bf 3.1. The model, main features}

The simplest form of charged matter is a distribution that consists of
one type of charged particles; in this case  
  the charge density is just $e\rho$, where $e$ is a unit
of charge. To introduce the radiation field we first include the radiation
action
$$
A_{Radiation} =    \int dx \sqrt{-g}\Bigg({-1- \sigma
\over 16\pi} g^{\mu\nu}g^{\lambda\rho} F_{\mu\lambda}F_{\nu\rho}
-W[\sigma] \Bigg) =:  \int dx \sqrt{-g}\,{\cal L}_{\rm rad}.
\eqno(3.1)
$$ 
Some justification for this expression, that differs from the Maxwell
action by the inclusion of a ``photon density" field $\sigma$ and an
internal photon energy
$W[\sigma]$, was offered in the Introduction. For the functional
$W[\sigma]$ a preliminary suggestion may be to take
$$
W[\sigma] \propto \sigma ^2;  
$$
this choice will be motivated below, with other possiblities, and
critically examined in Section 5. Note that the field strength represents
the global, coherent electromagnetic field; the photon gas appears only by
way of the density
$\sigma$.  We defer discussion of the physical or thermodynamic meaning
of this field.
 
  In addition we include  a coupling of the potential to the
conserved current. The following modified  matter action is
formally gauge invariant,
$$
A_{Matter} =  \int d^4x\sqrt{-g}\Bigg({\rho\over 2} (g^{\mu\nu}
\psi_{;\mu}\psi_{;\nu} - 1) - V[\rho]\Bigg),~~ \psi_{;\mu} =
\partial_\mu\psi +eA_\mu .\eqno(3.2)  
$$
The constant $e$ is a unit of charge. 
For the potential $V[\rho]$ we shall adopt a simple form that leads to
an isentropic equation of state; see below.  

In the non-relativistic limit, defined by setting $\psi = ct +
\Phi,~ \vec v =  -\vec\bigtriangledown\Phi$, 
$$
A_{Matter} = \rho_{cr}\int d^4x \Bigg( \rho  (\dot\Phi + e  A_0- \vec
\pi^2/2 
 ) - V[\rho]\Bigg),~~ \vec \pi = \vec v  - e  \vec  A.  
$$
Wave equations, besides the modified Maxwell equations, are
$$
\dot\Phi + e A_0 -  \vec \pi^2/2 = dV[\rho]/d\rho,~~ \dot \rho +
\vec\bigtriangledown(\rho   \vec\pi) = 0 
$$
As usual, define the matter pressure $p$ to  be the Lagrangian density 
or, equivalently, by the familiar formula
$$
p = \rho{dV\over d\rho}-V.\eqno(3.3) 
$$
The continuity equation shows that $ \vec
\pi$ must be interpreted as the velocity. The force balance equation
should be related to the gradient of the first wave equation, 
$$
 -(d/dt)v_i+e \partial_iA_0-  \vec\pi \cdot \partial_i\vec \pi =
\partial_i {dV\over d\rho} = {1\over\rho}\partial_i   p, 
$$ 
or
 $$
{d\over dt}   \pi_i +    \vec \pi \cdot \partial_i
\vec \pi   + {\partial_i  p\over  \rho} =
e \Big(F_{i0} + 
  (\vec \pi  \wedge B)_i\Big).\eqno(3.4)
$$
This agrees with standard theory of non-neutral plasmas [D] if we identify
the velocity with
$
\vec\pi$ 
   and the stress tensor with $\delta p  $ (= diag  $p$). We have not
found any discussion of a the possible presence of a photon gas in the
literature on nonneutral plasmas.

The energy momentum tensor has the two contributions, from matter and
radiation:
$$ 
T_{\mu\nu} =  \rho\psi_{;\mu}\psi_{;\nu} - g_{\mu\nu}p  
 + {-1-\sigma\over 4\pi}g^{\lambda\rho} 
F_{\mu\lambda}F_{\nu\rho} - g_{\mu\nu}\, \tilde {\cal L}.\eqno(3.5)
$$
\ve

\ce{\bf 3.2. The conserved current and the mass}

 The  conservation law (2.14) can be integrated to yield
$$
{d\over dt}\int_0^\infty\sqrt{\e^{(-\nu + \lambda)/2}} r^2
\rho(\dot\psi + eA) dr =
 \Big[\sqrt{\e^{(\nu-\lambda)/2}} r^2\rho(\psi'+eA_1)\Big]_0^\infty.
$$
 In view of the boundary conditions at the origin, 
$$
{d\over dt}\int_0^\infty\sqrt{\e^{(-\nu + \lambda)/2}}   \rho(\dot \psi
+eA_0)r^2dr  =
  \lim_{r\rightarrow \infty}\Big[\sqrt{\e^{(\nu-\lambda)/2}}
r^2\rho(\psi'+eA_r)\Big]. 
$$
The factor $\rho$ on the right hand side suggests that there is no flux at
infinity, but in fact the flux $ r\rho(\psi'+A_1)$ is   equal to
$-\dot \lambda/8\pi G$ by virtue of  Einstein's equations. 
For a static configuration both sides of the equation are zero; for a
first order deviation from a static configuration we have 
$$
{d\over dt}\int_0^\infty\sqrt{\e^{(-\nu + \lambda)/2}} r^2 \rho\delta
(\dot\psi + e A_0)
drd\Omega =  {1\over 2G}
 \lim_{r\rightarrow \infty} 
(r\delta\dot\lambda).
$$
If the perturbed and unperturbed metrics both tend  to Schwarzschild at
infinity, then
$r \dot\lambda \rightarrow 2 \dot mG$ and 
$$
{d\over dt}\int_0^R\sqrt{\e^{(-\nu + \lambda)/2}} r^2
\rho\,(\delta\dot\psi + e\delta A_0) drd\Omega = \dot m
. 
$$ 
It is not {\it a priori} obvious that the left side is  a constant
of the motion, but the result of our calculations  is that $r\delta\lambda$
and $r\delta A_0$ tend  to zero at infinity so that in fact $\dot m =
\dot q = 0$. The integral
$$
 \int_0^\infty\sqrt {-g}\,  g^{tt}
\rho  r^2(1+eA_0) dr d\Omega  
$$  
is a constant of the motion.

\b

\ce{\bf 3.3. Static configurations}

Since $ \vec\pi$ is interpreted as velocity, we may call `static'
a configuration in which $\vec\pi = 0$, $  \dot\psi = 1$ and $\dot \rho
= \dot A_0 = \dot\nu = \dot \lambda = 0$. There is a gauge in which $\vec
 A = 0$, Maxwell's equations reduce to Poisson's equation. With $A_0
\rightarrow A$, 
$$
 ( {\e}^{-(\nu + \lambda)/2}\, r^2 (1+\sigma) A')' =  
4\pi       
\rho e
\,{\e}^{(\lambda - \nu)/2}\,r^2(1 + e A),
\eqno(3.6)
$$ 
and variation of the field $\sigma$ gives
$$
 {-1\over 8  \pi} \, g^{tt}g^{rr} F_{tr}^2 =
{1\over 8\pi } \e^{- \nu -
\lambda } A'^2  = {d W\over d\sigma} 
 .\eqno(3.7)
$$  
  
Derivation of (3.7) with respect to the radial coordinate gives
$$
{1\over 8\pi } \e^{-\nu-\lambda}\Big( -(\nu + \lambda)' A'^2 +
2A' A''\Big) = {  p_{\rm ph}'\over \sigma},\eqno(3.8) 
$$
where 
$$
p_{\rm ph}  := \sigma{d\ W \over d \sigma} - W
$$
may be interpreted as the pressure of the photon gas. 
The first term on the left hand side is negative, which favors a pressure
that decreases with distance, just as is the case with ordinary matter. If
$A'^2$ decreases with distance then the second term is also negative,
which   represent an additional attractive force holding the photons
together. At a large distance this term predominates.
 
 Of the matter wave equations there remains only  the integrated
hydrostatic condition,
$$
  g^{00}(1 + e   A)^2 - 1  = 2{dV\over d\rho} = 2a\gamma f.\eqno(3.9)
$$
The Emden function $f$ is related to the density and the pressure
by                                 
$$
\rho = f^n,~~ p = {a\over n}f^{n+1},\eqno(3.10) 
$$
where $a$ is a constant that depends on the type of matter and that 
shall be regarded as a free parameter. Derivation of (3.9) with respect
to the radius gives
$$
\e^{-\nu}\Big(-\nu' (1 + e  A)^2 + 2e A'(1 + e A)\Big) =
{ p'\over\rho}.
$$
The first term is negative as usual. The second term is due to the
electrostatic interaction; since it is evidently repulsive  we conclude
that $e A'$ must be positive at large distances.  

We shall
look for static solutions with spherical symmetry, with the metric  
 (2.19),
$$
ds^2 = \e^\nu dt^2 -\e^\lambda dr^2 - r^2 d\Omega^2,
$$
the coefficients now depending on the coordinate $r$ only, $0<r<\infty$.
Einstein's equations reduce to 
$$\eqalign{ 
  1 -(r\e^{-\lambda})'&=  8\pi G  r^2\Big(
{\e}^{-\nu}\rho(1 + e A)^2  -p\Big)  +8\pi G r^2\Big(\e^{-
\nu - \lambda}{ 1+\sigma\over 4\pi} A'^2  -
   {\cal L}_{\rm rad}   \Big) ,\cr
\nu' + \lambda'   &=  8\pi  G\,    \rho\,r \e^{-\nu}(1+e
A)^2.   
  \cr}\eqno(3.11)
$$
They must be supplemented with (3.7) and (3.9). 
 
\b
\ce{\bf 3.4. Reissner-Nordstrom and infinite distributions} 

The exact
Reissner-Nordstrom solution of Einstein's equations in empty space has
two free parameters, the mass
$m$ and the charge $q$,
$$
 e^\nu = \e^{-\lambda} = 1 -  2mG /r + q^2 G  /r^2,~~  A  =
 -q/r.\eqno(3.12)  
$$
In the case that $\rho=  0$ our equation (3.9) gives
$ g_{00} = (1 + e A)^2$ which agrees with (3.12)  if   
 $  q^2 = m^2G$ and $e^2   =  G$; that is, in the case of the
extremal  Reissner-Nordstrom solution.  In this special case the
radiation pressure is balanced by the attractive force mentioned
after Eq.(3.8).  

If
$q^2> m^2G$ the electrostatic repulsion dominates and it is to be
expected that the only static solution 
 is a space that is empty of matter except
possibly near the center, where  gravity may become strong. In this case,
when there is no horizon,
we agree with the traditional interpretation of the Reissner-Nordstrom
metric as the metric of a space that is empty except for a small region
near the center.  If instead
$q^2<  m^2G$ we expect to find solutions with non zero matter density.
The metric may approach that of Reissner-Nordstrom at great distances but
it will have no horizon. The exact Reissner-Nordstrom solution with
$\rho  = 0 $ outside the horizon can perhaps be interpreted as a singular
limit of a sequence of space times with mass concentrated near and inside
the horizon.  

In general we expect that, if the density is integrable at infinity,
there are solutions that behave for large $r$, to leading order in $1/r$,
as the Reissner-Nordstrom solution. Assuming that there are numbers
$m,q $ and $\epsilon$  such that, asymptotically,
$$
\lambda - 1 \sim 1-\nu \sim {2m\over r},~~~ A \sim -{q\over r},~~ \rho
\sim   r^{-\epsilon }.
$$
we find from (3.10)  that $\epsilon \geq n$ and that
\b
$\bullet$ If $\epsilon > n$  , then $e  q =m G$.

$\bullet$ If $\epsilon = n$, then $mG-e  q = (2a/n)\gamma \lim rf$,
so that
$e  q < mG$.
\b
\no There are no solutions of this kind if $eq > mG$. There are no
solutions such that that $\rho = 0$ everywhere. Further
analysis of the asymptotics indicate that $\epsilon = n$ so that
$f\propto 1/r$.

\b

\ce{\bf 3.5. The photon equation of state}

Asymptotically, at large distances, the equations imply that the density
$\sigma$  is of the order of $1/r^4$.
Both $W$ and
$W'$ fall off as $1/r^4$, this gives internal evidence about the
functional $W[\sigma]$.  
Eq.(3.7) implies that 
$$
  W =  b~\sigma^2(1 +
b_1 \sigma + ...), 
$$
with $b,b_1,... ~$constant. This gives the pressure of the photon gas,
$$
  p_{\rm ph} = \sigma W' - W =b\,\sigma^2(1 + 2b_1 \sigma +
3b_2\sigma^2...).
$$
The contribution of the photon gas to the  ``energy density" $T_t^t$ is
$$ 
 {-\sigma\over 8 \pi}F^2-  p_{\rm ph} =   2\sigma W' -\sigma W' +W =
\sigma W' + W.
$$
If we retain only the first term in the expression for $W[\sigma]$ it is
$$
 \rho W' + W = 3b\,\sigma^2 = 3  p_{\rm ph},
$$ 
which is the familiar equation of state for a photon gas.
The most reasonable equation of state, both from contextual and
thermodynamic considerations is thus associated with the simple expression
$$ 
W = b\, \sigma^2,\eqno(3.13)
$$
where $b$, unlike $a$, is a fundamental constant, the value unknown to us
so far.

Further discussion of this choice of the functional $W[\sigma]$ is
deferred to the last section. 
\ve

\ce{\bf 3.6. Exact solutions} 

It is curious that the matterless Reissner-Nordstrom metric can be
derived from (3.9),
the equation derived from the action principle by variation of $\rho$,
in the special case that $\rho = 0$.  The complete set of equations is in
this cae, when $\rho = 0$,
$$\eqalign{
\nu + \lambda &= 0,\cr r^2(1+\sigma)A' &= q^2 = {\rm constant},\cr
A'^2 &= 16\pi b\sigma,\cr
1-(r\e^{-\lambda})' &= 8\pi Gr^2(2b\sigma + 3b\sigma^2), \cr}
$$
and, from variation of $\rho$,
$$
\e^\nu = (1 + eA)^2.
$$
The first 4  equations give us
$$
(1+\sigma)^2\sigma = (q^4/16 \pi b)r^{-4},
$$
and if $A' = \sqrt{64\pi b/3}~X$,
$$
(3 + 4X^2)X = \sqrt{27\over 64\pi b} ~ {q^2\over r^2} = c/r^2.
$$
If $\theta$ is defined by $X = \sinh\theta$, then the left side is $\sinh
3\theta$ and an explicit formula for $A'$ is obtaind in the form
$$
A' = \sqrt{64\pi b\over 3} \sinh({1\over 3}\sinh^{-1}{c\over r^2}).
$$ 
The metric is determined and  when it is eliminated we are left with the
following,
$$
1-\Big(r(1+eA)^2\Big)'= 8\pi Gr^2(\sigma W' + W +  W').
$$
The existence of a solution with no matter thus implies a special form of
the functional $W[\sigma]$. This solution gives $A'$ and $\sigma$ as
smooth, positive functions that fall of as $1/r^2$ and $1/r^4$,
respectively. Near the center,
$$
A \propto r^{1/3},~~\sigma\propto r^{-4/3},~~\lambda\propto r^{2/3}.
$$
This is not what is considered normal on physical grounds. The conclusion
is that the Reissner-Nordstrom solution is not rendered more physical by
the inclusion of a photon gas. Changing the equation of state does not
help. It remains to be seen whether the inclusion of matter leads to more
reasonable results. For this we must appeal to numerical solutions.

\ve
\no{\steptwo 4. Numerical calculations}

The full
set of equation in the static case consists of Einstein's equations,  
 $$\eqalign{  
\e^{-\lambda}(r\lambda' -1) + 1  = & 8\pi (G\rho_{cr})r^2
\Big(\e^{-\nu}f^n(1 + eA)^2  -(a/n)f^{n+1}\Big)\cr & + 8\pi G r^2\Big(
2b\sigma + 3b\sigma^2   
\Big),\cr}\eqno(4.1)
$$
$$
\nu' + \lambda'    =  
8\pi (G\rho_{cr}) r  \e^{-\nu}\rho \,(1+e A)^2,\eqno(4.2)
$$
Maxwell's equations,
$$  
 \big( {\e}^{-(\nu + \lambda)/2}\, r^2 (1+\sigma) eA'\big)'  =  4\pi \,
(e^2\rho_{cr}) 
\,   r^2\rho
\,{\e}^{(\lambda - \nu)/2} (1 + e A),\eqno(4.3) 
$$
and the equation that come from variation of the densities,
$$ 
 {1 \over 8\pi }\e^{-\nu-\lambda} A'^2 = W'   
 ,~~~  \e^{-\nu}(1 + e A)^2 - 1 =  2a\gamma f.\eqno(4.4-5)
$$
 
 We set  $b = \rho_{cr}\beta$, absorb
$e$ into $A$ and set  $\alpha = e^2/G$.  Replacing $(G\rho_{cr})$ by 1  in
the equations amounts to fixing the unit of length. 
  The final form of the equations is
 $$\eqalign{  
\e^{-\lambda}(r\lambda' -1) + 1  =  &8\pi   r^2 \Big(\e^{-\nu}f^n(1 +
A)^2  -(a/n)f^{n+1}\Big)\cr & + 8\pi  r^2\Big( 2\beta\sigma +
3\beta\sigma^2   
\Big),\cr}\eqno(4.1)
$$
$$
\nu' + \lambda'    =  
8\pi   r  \e^{-\nu}\rho \,(1+  A)^2,\eqno(4.2)
$$
$$  
 \big( {\e}^{-(\nu + \lambda)/2}\, r^2 (1+\sigma) A'\big)'  =  4\pi \alpha
\,  
\,   r^2\rho
\,{\e}^{(\lambda - \nu)/2} (1 +   A),\eqno(4.3) 
$$
$$ 
 {1 \over 8\pi }\e^{-\nu-\lambda} A'^2 = 2\alpha\beta\sigma   
 ,~~~  \e^{-\nu}(1 +   A)^2 - 1 =  2a\gamma f,\eqno(4.4-5)
$$
with $\alpha = e^2/G$.

As always, we assume regularity of all fields at the center. We start
from $\alpha =  \beta = 0$. In this case the electromagnetic sector
makes no contribution and we already have solutions for $(n_1,n_2) =
(3,6), a> 6$, given in the Table. We choose a value of $a$ at random,
$a = 1/10$, and begin to increase the values of $\alpha$ and $\beta$. All
the parameters are now fixed and we adjust the values of $\nu(0)$
and $A(0)$, with
$A'(0) = \lambda(0) =0$, until we get a solution for the function $\nu$
such that
$2m = \lim r\nu$ at infinity is finite. The idea is to reach the extremal
value
$\alpha = e^2/G = 1$, adjusting $\beta$ and $a$ as necessary as we
gradually increase $\alpha$ from zero. The value of the parameter $\beta$
turns out to be nearly irrelevant. More precisely, a rescaling of $\beta$
does nothing more than rescale the solution for the field $b$ by the
inverse factor.

The search of solutions that exist for a discrete set of points in a
2-dimensional parameter space is very laborious. Representative solutions
are given in the Table. The highest value that was attained for the
Reissner-Nordstrom parameter $\alpha = e^2/G$ was .65, but this is
probably not an absolute limit. 

\ve
\ce{\bf Table}   \font\stephalf=  cmr6
 scaled\magstep0

{
 \settabs \+   &~~~~~~~\quad 
&~\quad~\quad~~~~~&\quad~~~~~~~~~~\quad\quad &\quad~~\quad\quad
&\quad\quad~~~~~\quad&~\quad\quad\quad
&~ \quad\quad~~  \quad&~~~~~~`~\quad&\quad~~\cr
\+&$a={1\over 6}$&$b=1$& $-\nu(0)$&$~f(0)$&$-A(0)$&$~R$&$2mG$&$q$
& maximum \cr  
\+& &$k= 0$&.5670415&1.749&& .0962&.03045& \cr
\+& &$k= .15$&.61132&1.75&.03221&.10&.03935&.00256&$\sigma(.10)=.0052$ \cr
\+&&$k = .40$
&.699495&1.42&.0975216&~.115&.0654&.00995&$\sigma(.13)=.0123$&\cr
\+&$a ={1\over 10}$&$k = .45$\cr

\+&&$b = .10$&.448622 &1.30&.0729404&.10&.0439&.0079&$\sigma(.12)= .27$\cr
\+  && $b=1.0$&.443985 &1.21&.0793518&.10&.04838&.00868&$\sigma(.125) =
.035$\cr
\+&& $b = 10$&.44334&1.20&.080233 &.10 &.0491&.0088 &$\sigma(.130)=
.0035$&
\cr
\+&$a={1\over 20}$&$b=1$ \cr
\+&&$k = 0$&.1471706&1.20&&.065&.00886& \cr 
\+&&$k = .695$&.32936&1.02&.10005136&.04&.0607&.0167&$\sigma(.125)= .016$
\cr}

\b
\no{\bf Table.} The first group bshows the effect of increasing the
elementary charge, $k = e^2/G$. The second group shows that a change in
the strength $b$ of the free energy affects only the density $\sigma$.
The last entry give the highest value of $k$ for which a solution was
found.
 
 
\b
\ce{\steptwo 5.  Discussion}

\ce{\bf 5.1. The photon gas}

There seems to have been no previous discussion of the role of the photon
gas in a non neutral plasma. The need for an action principle is 
very keenly felt. The inspiration for the specific action (3.1) is the
form of the action (1.1) that has been successful in the treatment of the
matter component. The appropriateness of treating the global
electromagnetic field as a complement of the photon gas
(the inclusion of the constant term in the coefficient of $F^2$ in the
action)  is a guess that must be evaluated {\it a posteriori}.

Our initial choice for the functional $ W[\sigma]$, 
$$
W[\sigma] = b \sigma^2,~~ b ~{\rm constant},
$$ 
is motivated within the context by simplicity and asymptotics. As has
been stressed elsewhere, our program is to study the  coupling of the
gravitational field to interesting field theoretic models via an action
principle. The physical interpretation has to come from the model itself
and a  confrontation with classical thermodynamics is not guaranteed to
give results that are entirely as expected, especially in the case of
strong fields [F2]. But a difficulty arises when it comes to choosing the
detailed form of the action. In this paper we are testing an
 action  of the form  (3.1), parameterized by the choice of
$W[\sigma]$, for a particular choice of this functional, and it is
interesting to ask what are the thermodynamic  implications. Bluntly:
what choice of
$W[\sigma]$ is favored by   classical thermodynamics?

As was already pointed out, the choice $W[\sigma] =b \sigma^2$ gives
the familiar relation between the energy density and the pressure of a
pure photon gas. The association of $\sigma$ with $F^2$, Eq.(3.7), allows
us to interpret the on shell value of the potential as being proportinal
to
$F^4$, a Born-Infeld modifier that can be attributed to the scattering of
light by light, which determines the numerical value of the constant $b$.
(Since
$b$ is a very large, the photon density will normally be very small.)
This photon self-interaction is the physical origin of the photon
pressure, allowing the photon gas to transmit sound at velocity
$c/\sqrt 3$.

The inclusion of charge, and the need for an electrostatic field of the
form $q/r$, forces us to include the term $-F^2$ in the action density.
An implication of this is that the ratio (total energy density)/(total
pressure) is now longer equal to 3 in the presence of charged matter. If
this conclusion is correct then we should not expect this last relation
to hold except in the case of a pure photon gas, with no matter present,
whether charged or not.   

Each equation of state, for matter and for the photon gas,  can be 
characterized as isentropic. If the matter component behaves as an ideal
gas, then the temperature is proportional to $p/\rho$ and thus to the
Emden function, $f\propto T$. Asymptotically, $f $ falls off as $1/r$;
hence $T \propto 1/r$. If the photon gas behaves as expected by the
thermodynamic interpretation,  then the energy density, in the absence of
matter,  should be proportional to the fourth power of the temperature,
hence
$\sigma
\propto T^2$.   In a star made up of charged matter
and photons, both $\sigma$ and the photon energy density fall  off
as
$1/r^4$, and this component of the energy density is not 3 times the
corresponding component of the pressure. 

Strict thermodynamic equilibrium
would require that both temperatures be the same; hence $\sigma$ should
fall off as $1/r^4$, as predicted by Eq.(3.7).  The model is thus
internally consistent with a thermodynamic equilibrium at large distances
in the sense that the two temperatures are at least proportinal and
perhaps, with some fine tuning, equal. Near the center things are very
different, for the Emden function is approximately flat and non zero,
while $\sigma$ tends to zero. However, the naive expectation that each
component of the mixture behaves as if it were alone (and in the absence
of gravity)  was already discredited in the preceding paragraph. Probably
one should define equilibrium in mechanical terms, as a static solution,
and attempt to find a good definition of temperature for the mixture.

The most spectacular result of the numerical calculations, with
$W[\sigma] \propto \sigma^2$, is the fact that the density $\sigma$ of the
photon gas turns out to have a profile that is very different from that
of the matter density.  This is not consistent with the practice of
postulating that the two pressure profiles are proportional to each other.
(The traditional approach does not easily accomodate difference in
pressure.) In the case of a non neutral plasma they seem to be greatly
different.

Lastly, it may be pointed out that the traditional approach also
falls short of incorporating all the expectations based on
thermodynamics, even in the case of a neutral gas. It is usual to assume
that the pressures of radiation and of matter are proportional. For  a
perfect gas it is 
 $\rho = f^{n+1} \propto T^{n+1}$, so  this comes
out right only in a  region where $n = 3$.

\b

\ce{\bf 2. Suggestions}

The action (3.1), originally inspired by analogy with the matter action
(1.1) has turned out to incorporate some if not all of our experience in
dealing with electromagnetic phenomena. Our attempts to justify the
specific form of this action has led to a partial understanding of the
interpretation of the density $\sigma$. Indeed, it can be seen as an
effective representative of the dielectric properties of the photon gas.
In the case studied here, in the absence of flow and of magnetic fields,
$$
\sigma F^2 = -\sigma \vec E^2 = -\vec D\cdot \vec E.
$$
Interactions between the electromagnetic field and charged matter may be
adequately described by the coupling to the conserved current, but it
seems plausible that dielectric properties of the matter gas may have to
be taken into account separately. In the case of neutral matter this
becomes dominant and it may be thought that the main source of
matter-field interaction may be the inclusion of a dielectric
modification of Maxwell's action:
$$
-\int dx\sqrt{-g}(1+ \epsilon)F^2,
$$
where the field $\epsilon$ is plausibly taken to be proportional to the
density $\rho$.

The reality of the photon gas strongly suggests that there may be
circumstances in which an analogous graviton gas may become interesting.
Indeed, this suggestion has been made in the context of stars -
gravistars [MM] - as well as an effect to be taken into account in
cosmology [R].

F. de Felice, Yu Yunqiang and Jing Fang : Relativistic charged spheres
Mont. Not. R. astron. Soc London  277 (1995) L17

F. de Felice, Yu Yunqiang and Liu Shiming: Relativistic charged spheres: Regularity  and stability
Class. Quantum Grav. 16 (1999) 2669

\b

\no{\steptwo Acknowledgements}

Useful conversations with R.J. Finkelstein, R.W. Huff, W. Mori and S.
Putterman are gratefully ackonowledged.
\b\b
\no{\steptwo References}

\no[dF1]~ de Felice, F., Yu Yunqiang and Jing Fang, Relativistic charged
spheres 

			~~~Mont. Not. R. astron. Soc London  277 (1995) L17.

\no[dF2] ~de Felice, F., Yu Yunqiang and Liu Shiming, Relativistic
charged spheres:  

	~~~Regularity and stability, Class. Quantum Grav. 16 (1999) 2669.

\no [F1]  ~~\hskip.8mmFr\o nsdal,  C.,  Ideal Stars and General
Relativity, gr-qc/0606027.

\no [F2]  ~ \hskip.8mm Fr\o nsdal, C., Stability of polytropes, arXiv
0705.0774 [gr-cc]

\no [D]~~ ~ Davidson, R.D., ``Theory of Non-Neutral Plasmas",
Addison-Wesley
  1990.

\no [MM] ~Mazur, P. and Mottola, E., Gravitational Vacuum Condensate
Stars, 
 
\no \quad \quad~~~gr-qc/0407075     
 
\no [R] ~~ ~Rees, M.F., Effects of very long wavelength primordial
gravitational radiation, 

\quad    Mon.Not.astr.Soc. {\bf 154} 187-195 (1971).

\end
   \font\stephalf=  cmr6
 scaled\magstep0

\ce{\bf Table 1 .}
{\stephalf
 \settabs \+   &  ~~~~~~\quad 
&~\quad~\quad~~~~~~&\quad~~~~~~~~~~~~~\quad\quad &\quad~~~~\quad\quad
&\quad\quad~~\quad&~~~\quad\quad\quad & 
 \cr
\+ &$e$&
$~~~-\nu(0)$&$-A(0)$&$~f(0)$&$~~q$&$~2m$&$~~R$\cr                   
\+ & 0&.1765827&~~0&.0724 &~~0&~~.78&~~10.4\cr
\+&.1&.1780627&.00795345&.0724&~~.0395&~~.792&~10.2\cr
\+&.5&.22055865 &.0488269&.07&~~.278&~1.22&~11.5\cr
\+&.7&.280702&.08609456&.0635&~~.634&~2.205&~15.0\cr
\+&.8& .332585&.11568445 &.0556&~1.11&~3.69&~20\cr
\+  &.9&.4311& .16580614 &.0427&~2.8&~9.40&~38&\cr
\+&.91& .4505&.17440770195&.0412&~3.3  &11.12 &~43 && \cr
\+&.92&.47426&.184570355&.04&~4.05&13.7&~52\cr
\+&.93&.50370&.1967569707&.0392&~5.25& 17.7&~66& \cr
\+&.935& .5200&.20336678&.0387&~6.1&20.2&~76\cr
\+&.94&.5400&.211318832&.0383&~7.3& 25.0&~88& \cr
\+&.945&.5602&.2192118485&.0378&~8.5&30.5&105\cr
\+&.95&.5804&.2210&.0373&11.6&39.3&129\cr}

\b\b\ce{\bf Table 2 .}
{\stephalf
 \settabs \+   &  ~~~~~~\quad 
&~\quad~\quad~~~~~~~&\quad~~~~~~~~~~\quad\quad &\quad~~\quad\quad
&\quad\quad~~\quad&~~~\quad\quad\quad & 
 \cr
\+ &$e$&
$~~~-\nu(0)$&$-A(0)$&$ f(0)$&$~~q$&$~~~2m$&$~~R$\cr                   
\+ & 0&1.064561&~~0&.712 &~~0&~~.3385&~4.5\cr
\+&.1&1.072838&.0293768&.715&~~.0156&~~.343&~4.45\cr
\+&.5&1.3287795 &.1678605&.813&~~.089&~~.460&~4.8\cr
\+&.7&1.711247&.2674352&.996&~~.134&~~.582&~5.0\cr
\+&.8&2.0161155&.32649472 &.1.15&~.151&~~.641&~4.9\cr
\+  &.82&2.088306& .3387496 &1.20&~.151&~~.650&~4.9&\cr
\+&.86& 2.244881&.3634264&1.30&~.158  &~~.666&~4.85 && \cr
\+&.8642&2.2622&.36602&1.308&~.1583&~~.667&~4.85\cr

 \font\stephalf=  cmr6 scaled\magstep0  
 
\b\b\ce{\bf Table 1. Limits on the parameter $a$ in the case $e = .1,
~ \tau = 1/30,~ n = (3,6)$ }
 \def\s{\scriptstyle}{\stephalf
~~
\settabs \+ ~~~~~~~~~~~~~~~~~~~~~~     & 1~~~~~  ~ ~  ~~~~& 20~~~~~~~~
~ ~& 100~~~~~~~~~~& 1000~~~~~~~& 10000~~~~&~~~~~~~~~~ &
100000~~~~&~~~~~~~~~~~~~&~~~~~\~~~~~~~&~~~~~~~~~~~~&~~~~~~~~~~~~&~~~~~~~~~~~~~&~~~~~~~~~\cr
 \+&& - $\s\nu$(0)  
  &  -A(0) &f(0)& R&2m&q&f(R) &\cr
\+&a = 1.58 &.971336&.055882&.383&29&3.37&.31 &.033\cr
\+&a= .543&1.2409&.0633071&1.67&19.5&.85&.08&.033&\cr 
\+&...\cr
\+&a = 2.591&.68463&.0455675&.14&48&8.64&.81&.034\cr 
\+&a = .3394&.0593400&.0057137&.0662&34&.95&.092&.034\cr
\+&...\cr
\+&a = .86&1.63932222&.071312&1.78&29&1.995&.185&.034\cr 
\+&a = .68&1.463837&.06817365&1.80&26&1.38&.13& .033\cr}

\b
Table 1. When the elctric potential is included it is no longer true
that the value of the pressure parameter $a$ merely sets the scale.
The table shows the range of allowed values of $a$ for three
points in Fig.1  that have the critical densith, $\tau = 1/30$.

\b
For our next project we fix $a = 1,~\tau = 1/30$ and try to increase
the unit of charge in the hope of getting close to the extremal
Reissner-Nordstrom solution in the exterior. The radius remains
fixed at $R \approx 11$ as we increase the charge unit $e$.
The highest value of $e$ for which a solution was found is .8106.
Table 2 shows the variation of the other parameters.
\b
\ce{\bf Table 2. The effect of increasing the unit of charge}
\b
\settabs \+ ~     & 1~~~~~  ~ ~  ~~~~& 20~~~~~~~~ ~ ~& 100~~~~~~~~~~&
1000~~~~~~~& 10000~~~~&~~~~~~~~~~ &
100000~~~~&~~~~~~~~~~~~~&~~~~~\~~~~~~~&~~~~~~~~~~~~&~~~~~~~~~~~~&~~~~~~~~~~~~~&~~~~~~~~~\cr
 \+&&e(0)  
  &$\s -\nu$(0)&  -A(0) &2m& q& \cr
\+& &0&.1765829&0& .78& 0 & \cr
\+& &.1&.1783625&.0079658&.78&.04& \cr 
\+& &.2&.18393& .0164034&.82& .075 & \cr
\+& &.5&.2359085&.051850&1.11&.276& \cr 
\+& &.7&.35481&.10555595&1.705& .60 & \cr
\+& &.8&.56542&.1800899&2.22&.91& \cr
\+& &.81&.628354& .1982669&2.2& .92 & \cr
\+& &.814&.6776&.21099224&2.138&.899& \cr

\end 
 
\ce {\bf 3.4. Polytropes with boundary}

Equations (3.7,9,10) can be solved numerically in the case of a
homogeneeous and isentropic equation of state.

Suppose that $V[\rho] = a\rho^\gamma,~ a~$constant. The equation of
state is that of the polytrope with index  $n = (\gamma-1)^{-1}$,
$$
p = {a\over n}\rho^\gamma.
$$
The constant  $a$ will be treated as an adjustable parameter. The scale
transformation
$$
x = r\sqrt{8\pi Ga^{-n}},~~ a^n\rho(r) =\bar \rho(x),~~ el = \bar e \sqrt
G, ~~ A_0 = \bar A_o/\sqrt G, 
$$
reduces Eq.s (3.7-10)
to
$$\eqalign{&
 (x^2 {\e}^{(\nu+\lambda)/2}\bar A_0')' = {\bar e \over
2 }x^2\,{\e}^{(\nu+\lambda)/2}\rho\,(1 + \bar e \bar A_0), \cr &
{\e}^{-\nu} (1 + \bar e \bar A_0)^2 = 1+2\gamma \rho^{1/n}, \cr &
{\e}^{-\lambda} (x\lambda' -1) + 1  =  
x^2\big( {\e}^{-\nu}\rho(1 + \bar e\bar A_0)^2  -\bar p  + 
 {\e}^{-\nu-\lambda}\bar A_0'^2\big),\cr &   
{\e}^{-\lambda}(-x\nu' -1) + 1  = x^2\big( -\bar p+ 
 {\e}^{-\nu-\lambda}\bar A_0'^2
 \big), 
\cr}
$$
with $\bar p = \bar \rho^\gamma/n$. Henceforth we drop the bars as well
as the suffix.   Then
$$\eqalign{&
 (x^2 {\e}^{(\nu+\lambda)/2}A')' = {e \over
2 }x^2\,{\e}^{(\nu+\lambda)/2}\rho\,(1 +  eA), \cr &
{\e}^{-\nu} (1 +  e A)^2 = 1+2 \gamma \rho^{1/n}, \cr &
{\e}^{-\lambda}(x\lambda' -1) + 1  =  
x^2\big( \rho + {n+2\over n}p  + 
 {\e}^{-\nu-\lambda} A'^2\big),\cr &   
\nu' + \lambda'  = x^2{\e}^\lambda\big(\rho + 2{n+1\over n} p)  
\cr}.
$$
If $\rho$ is integrable at infinity, then for large $x$
there are constants
$m,q$ such that
$$
\lambda - 1 \sim 1-\nu \sim {2m\over x},~~~ A_0 \sim {q\over x}.
$$
In the case that there is no radiation field, if $n<5$ and $q = 0$  there
is a family of solution that are regular at
$x = 0$, with $\nu'$ and $\lambda'$  vanishing there. For such
solutions there is a value   $X_0 $ of $x$  where the
pressure and the density pass through zero.  The boundary of the star is
therefore at some value  $  X<X_0$ of $x$ . At
this point we can match the metric   to an
exterior, Schwarzschild solution. This implies that
$$
\nu(X) + \lambda(X) = 0.
$$
When a solution is constructed by integrating outward from the center,
then this relation determines the value of $X$. For a fixed
polytropic index $n$ there emerges a relation between
$X$ and $m$ that is shown in a log-log plot in Fig.1, for the case $n =
3$, by the curve that is marked ``$e = 1"$.

We now expect to find solutions where not only the metric, but also the
field $A$ is regular at the origin. This would imply that
$A'(0) = 0$ and introduce a new parameter $A(0)$. We would increase
the value of this parameter from zero and observe the effect on the 
relation between the parameters $m$ and $R$ shown by the shorter curve
in Fig.1 for the case that $A(0) = 0$.

However, concerns about  the accuracy of the calculation for very small
values of $x$ promted us to integrate the equations by a power series in
$x$ in some small interval where 2 or 3 terms are sufficient.
Thus
$$ 
\nu(x) = a + b x^2,~~ \lambda(x) = cx^2,~~ A(x) = \alpha x^2.
$$
For small values of $\nu(0)$ the exponential functions can be replaced
by unity and the four equations yield, in the same order,
$$\eqalign{&
6 \alpha =e f^n/2,\cr &
\e^{-a} -1   =  2\gamma f,\cr &
3c= f^n  + {n+2\over n}f^{n+1},\cr &
2b+2c = f^n + {2n+2\over n}f^{n+1}.
\cr}
$$
Thus $f$, defined by $f^n = \rho(0)$, turns out to be the only free
parameter, besides the coupling constant $e$. The field $A$ must vanish
at the center. 

We thus have to accept the conclusion that, under the assumption of
regularity at the center, the charge $q$, if not equal to zero,  is
determined along with the mass and the radius by $e$ and  the value of
the function $/nu$ at the center. But to find the value of $q$ we have to
integrate the equations.

We have done so for the case that $n = 3$. Thus for small $x$ we used
$$\eqalign{&
A(x) = e(f^3/12)x^2,~~ \nu(x) = a + b x^2,~~ \lambda(x) = cx^2,
\cr & c = f^3 + (5/3)f^4,~~ b = (1/2)f^3 + (4/3)f^4 - c,~~ f =
(3/8)(e^{-a}-1).   
\cr}
$$
We used the same values of $a = \nu(0)$ as in the calculations that
produced the data in the neutral case. For the point at the right end $a =
.0001$. Increasing $e$ from $10^{-10}$ we find that deviations begin to be
appreciable as $e $ exceeds unity.
\b

\ce {\steptwo 5. Discussion of results}

\end
\ce{\bf Table 1, $n$ = 3}
 
\settabs \+ ~ & x~~~~~~~~~& 10~~~~~~ ~~  ~~~~& 20~~~~~~~~~   ~~~&
100~~~~~~~~~~~& 1000~~~~~~~~~& 10000~~~~~~~~&
100000~~~~~&~~~~~~~~~~~~~~~&~~~~~~~~~~~~~~~&\cr

 \+& $\nu$(0) = .0001&&&&& \cr 

\+& $e$  &.1&1&2&5&10&20&50&  \cr 

 \+& X&94 000 &94 000&94 000&94 000&94 000&93 000&90 000  \cr

\+& 2m&5.10 &5.10&5.10&5.10&5.10&5.10&5.10  \cr 

\+& $q$  &.215&2.15&  &10.8 &21.5& &109  \cr 

\b\hrule\b

 \+& $\nu$(0) = .1&&&&& \cr 

\+& $e$  &.1&1&2&5&10&20&50&  \cr 

 \+& X&85 &85&&62&44&28&14.5  \cr

\+& 2m&4.37&4.35&&3.93&3.31&3.16&1.67  \cr

\+ &q&.176&1.77&&4.67&4.78&3.16&1.67&\cr 
\b\hrule\b 

 \+& $\nu$(0) = .5&&&&& \cr 

\+& $e$  &.1&1&2&5&10&20&50&  \cr 

 \+& X&14&12&&6.0&3.6&2.16& 1.09 \cr

\+& 2m& &&&&&&  \cr

\+&q&.118&.90&1.29&1.06&.71&.55&.213&\cr
 
\b\hrule\b 

 \+& $\nu$(0) = 1&&&&& \cr 

\+& $e$  &.1&1&2&5&10&20&50&  \cr 

 \+& X&14.6&7.8&3.8&1.57&.9&.52&.255&  \cr

\+&LogX&2.68&2.054&1.335&.451&-.106&-.654&-1.366\cr

\+&2m&1.648&1.415&1.135&.732&.488&.307&.158\cr

\+&Log2m&.500&.347&.127&-.312&-.717&-1.181&-1.845\cr

\+&q&.243&1.074&.73&.347&.203&.110&.050\cr 
 \b\hrule\b 
\b 
We have plotted, also in Fig.1, by the longer curve, the relation between
$X$ and $2m$ for the case that $e = 5$.
 \ve
\ce{\bf Table 2, $e$ = 5}
\settabs \+ ~ & x~~~~~~~~~& 10~~~~~~ ~~  ~~~~& 20~~~~~~~~~   ~~~&
100~~~~~~~~~~~& 1000   & 10000~~ ~~&
100000~~~~~&~~~~~~~~~~~~~~~&~~~~~~~~~~~~~~~&\cr

\+& $\nu$(0)  &    X&2m&q&  &   $\nu$(0)  &    X&2m&q& &  \cr

 \+&.001   &94 000&5.08&10.8&&     1.2   &1.10&.541&.261&      \cr

\+&.005   &1810&5.03&9.97&&   1.3   &.92&.468&.220&&  \cr

\+&.02   &420&4.82&8.63&&   1.5   &.75&.364&.201&&  \cr

\+&.1   &62&3.93&4.67&&   1.6   &.75&.330&.223&&  \cr

\+&.3   &13.4&2.53&1.98&&  1.7&.925&.313&.323&\cr 

\+&.4   &8.5&2.08&1.43&&  1.8&3.30&.348&1.52&\cr

\+&.5&4.90&1.60&.66&&1.85\cr

\+&.8   &2.45&1.012&.511&&  1.9&23.5&.609&11.9&\cr

\+&.9   &1.95&.859&.422&&  2&14.0&.702&7.10&\cr

\+&1   &1.57&.732&.347 &&  2.05&11.2&.714&5.66&\cr
\b\hrule\b
\ce{\bf Table 3, $e$ = 5, logarithms}
\settabs \+ ~ & x~~~~~~~~~& 10~~~~~~ ~~  ~~~~& 20~~~~~~~~~   ~~~&
100~~~~~~~~~~~& 1000   & 10000~~ ~~&
100000~~~~~&~~~~~~~~~~~~~~~&~~~~~~~~~~~~~~~&\cr

\+& $\nu$(0)  &    X&2m&q&  &   $\nu$(0)  &    X&2m&q& &  \cr

 \+&.001   &11.45&1.625&10.8&&     1.2   &.095&-.614&.261&      \cr

\+&.005   &7.5&1.615&9.97&&   1.3   &-.083&-.759&.220&&  \cr

\+&.02   &6.04&1.57&8.63&&   1.5   &-.288&-1.01&.201&&  \cr

\+&.1   &4.127&1.369&4.67&&   1.6   &-.288&-1.109&-.288&&  \cr

\+&.3   &2.595&.928&1.98&&  1.7&-.078&-1.162&.323&\cr 

\+&.4   &2.14&.732&1.43&&  1.8&1.194&-1.056&1.52&\cr

\+&.5&1.59&.47&&&1.85\cr

\+&.8   &.896&.0012&.511&&  1.9&3.157&-.496&11.9&\cr

\+&.9   &.668&-.152&.422&&  2&2.639&-,354&7.10&\cr

\+&1   &.451&-.312&.347 &&  2.05&2.416&-.337&5.66&\cr

\ve
 
\ce{\bf Table 3, $e$ = 0}
\settabs \+ ~ & x~~~~~~~~~& 10~~~~~~ ~~  ~~~~& 20~~~~~~~~~   ~~~&
100~~~~~~~~~~~& 1000   & 10000~~ ~~&
100000~~~~~&~~~~~~~~~~~~~~~&~~~~~~~~~~~~~~~&\cr

\+& $\nu$(0)  &    X&2m&q&  &   $\nu$(0)  &    X&2m&q& &  \cr

 \+&.001   &94 000&5.096&10.8&&     1.2   &39.4&2.03&.261&      \cr

\+&.005   &1875&5.064&9.97&&   1.3   &45.07&2.26&.220&&  \cr

\+&.02   &463&4.97&8.63&&   1.5   &38.3&2.49&.201&&  \cr

\+&.1   &97&3.93&4.49&&   1.6    \cr

\+&.3   &25.2&3.50&1.98&&  1.7& \cr 

\+&.4   &18.0&2.08&3.10&&  1.8& \cr

\+&.5   &14.28&2.74&&&1.85\cr

\+&.8   &9.50 &1.99&.511&&  1.9& \cr

\+&.9   &10.84&1.83&.422&&  2&24.16&2.415&7.10&\cr

\+&1   &14.6&1.75&.347 &&  2.065&2.385&.714&5.66&\cr
\b\hrule\b

\ce{\bf Table 4, $e$ = 0, logarithms }
\settabs \+ ~ & x~~~~~~~~~& 10~~~~~~ ~~  ~~~~& 20~~~~~~~~~   ~~~&
100~~~~~~~~~~~& 1000   & 10000~~ ~~&
100000~~~~~&~~~~~~~~~~~~~~~&~~~~~~~~~~~~~~~&\cr

\+& $\nu$(0)  &    X&2m&q&  &   $\nu$(0)  &    X&2m&q& &  \cr

 \+&.001   &11.45&1.628&10.8&&     1.2   &3.67&.708&.261&      \cr

\+&.005   &7.54&1.622&9.97&&   1.3   &3.81&.815&.220&&  \cr

\+&.02   &6.13&1.603&8.63&&   1.5   &3.65&.912&.201&&  \cr

\+&.1   &4.66&1.502&4.67&&   1.6   & &&  \cr

\+&.3   &3.23&1.253&1.98&&  1.7& &\cr 

\+&.4   &2.89&1.13&1.43&&  1.8& &\cr

\+&.5   &2.658&1.008&&&1.85\cr

\+&.8   &2.25&.688&.511&&  1.9& &\cr

\+&.9   &2.38&.604&.422&&  2&3.185&,.882354&7.10&\cr

\+&1   &2.68&.560&.347 &&  2.05&.869&.337&5.66&\cr

\b
\no{\steptwo 4. Stability and radiation}

The homogeneous, charged polytrope is the simplest model of a star that
allows a study of radiation, and hence an investigation of the relation
between luminosity and temperature.

\end
\no{\steptwo References}

[D] Davidson, R.D., ``Theory of Non-Neutral Plasmas", Addison-Wesley
Redwood City 1990.

\no(1)
$$
\e^\nu = e^{-\lambda} = (1 + eA)^2,
$$
(2)
$$ 
{1\over 8\pi}A'^2 = W' := dW/d\sigma,
$$
(3)
$$
(r^2\sigma A')' = 0 ,
$$
and (4)
$$
1-(r\e^{-\lambda})' = 8\pi G \sigma W' + 8\pi G r^2W.
$$ 
Comments: We shall assume that
$$
A = {-q\over r} + o(1/r^2)
$$
at large distances. Then the third equation tells us that $\sigma$ must
have a finite limit; we choose this limit to be unity, then
$$
r^2\sigma A' = q.   
$$
The second and the fourth equation imply that both $W$ and $W'$ must
vanish when $\sigma = 1$.

Since $\sigma$ is not a constant (second equation), we must have to next
leading order (third equation),
$$
A = {-q\over r}(1- \alpha/r^n),~~ \sigma = 1+\alpha/r^n 
$$
We eliminate the metric and obtain from (4)
$$
{e^2q^2\over r^2} + {neq\alpha\over r^{n+1}} -{2(n+1)e^2q^2\alpha\over
r^{n+2}} = {Gq^2\over r^2}(1-{(n+1)\alpha\over r^n}) + 8\pi Gr^2W.
$$
So we have 2 conditions for $W$,
$$
8 \pi W' = {q^2\over r^4}(1-{2(n+1)\alpha\over r^n}),
$$
 
$$
8\pi W = {(e^2-G)q^2\over r^2} + {neq\alpha\over r^{n+1}}
-{2(n+1)e^2q^2\alpha\over r^{n+2}} +  {(n+1)Gq^2\alpha\over
r^{n+2}} .
$$
Normally, for photons,  one postulates the equation of state $p = W/3$,
which implies that $W[\sigma] \propto \sigma^{4/3}$. Thus
 $W'\propto \sigma^{1/3}$ If $ W'\propto 1/r^4$ then $\sigma\propto
1/r^{12}$ and $W \propto 1/r^{16}$.

Actually, if there is no matter present, one does not expect a potential
that falls off as $1/r$. Let us try instead
$$
A = {-q\over r}\e^{-\epsilon r}, ~~A' = q({1\over r^2} + {\epsilon\over
r})\e^{-\epsilon r},
$$
but this will make $\sigma$ increase exponentially at infinity.
The leading term in $W'$ is then of order $1/r^2$ and the left side of (4)
becomes
$$
-\Big((-eq\e^{-\epsilon r} + {e^2q^2\over r})\e^{-\epsilon r}\Big)'
= \Big(-eq\epsilon + {e^2q^2 \over r^2}    + {e^2q^2\epsilon\over
r}\Big)\e^{-\epsilon r}

\ve
 If $n = 1$ the first condition implies that $W'\propto (\sigma-1)^4$;
This suggests that $W\propto  (\sigma-1)^5\propto 1/r^5$, which is very
unexpected.  If $n = 2$ we get $W'\propto (\sigma-1)^2$, $W\propto
(\sigma-1)^3\propto 1/r^6$ and this cannot account for the second term,
of order $1/r^3$.